# Synchronization dynamics on the picosecond timescale in coupled Josephson junction neurons

K. Segall[1], M. LeGro[1], S. Kaplan[2], O. Svitelskiy[1], S. Khadka[1], P. Crotty[1] and D. Schult[3]

[1] Department of Physics and Astronomy, Colgate University, 13 Oak Drive, Hamilton, NY 13346
[2] Consultant, 1800 Cherokee Drive, Estes Park CO, 80517
[3] Department of Mathematics, Colgate University, 13 Oak Drive, Hamilton, NY 13346

Abstract: Conventional digital computation is rapidly approaching physical limits for speed and energy dissipation. Here we fabricate and test a simple neuromorphic circuit that models neuronal somas, axons and synapses with superconducting Josephson junctions. The circuit models two mutually coupled excitatory neurons. In some regions of parameter space the neurons are desynchronized. In others, the Josephson neurons synchronize in one of two states, in-phase or anti-phase. An experimental alteration of the delay and strength of the connecting synapses can toggle the system back and forth in a phase-flip bifurcation. Firing synchronization states are calculated >70,000 times faster than conventional digital approaches. With their speed and low energy dissipation ($10^{-17}$ Joules/spike), this set of proof-of-concept experiments establishes Josephson junction neurons as a viable approach for improvements in neuronal computation as well as applications in neuromorphic computing.

Introduction

The collective behavior of neural systems is a highly active area of study. It addresses basic questions from two major scientific challenges: understanding the human brain and building an efficient artificial learning processor. These questions require interdisciplinary examination, and even so are hard to make progress on. One approach takes a bottom-up approach, building and studying neural circuits at the cellular level. Another approach works top-down, looking to understand and emulate behaviors like synchronization,[1] information processing and memory.

Model systems are a useful tool to go between top-down and bottom-up approaches. One can simulate neuron behaviors in a network of cellular-level objects. Digital simulations can model the time-dependent membrane potentials of large numbers of neurons, but are often limited by computational time. Analog models, such as electrical Very Large Scale Integration (VLSI) circuits, sacrifice a certain amount of detail in their simulation capability, but can simulate the interactions in parallel, much faster than digital simulations. In fact, analog models have the capability to go faster[2,3] than the biology, potentially enabling studies of long-term behavior, learning and various neurological disorders.

In addition to providing a better understanding of neuroscience, analog model systems can be used to create artificial learning devices. Analog and digital neurons can be networked to form new kinds of "neuromorphic" processors that will help process complex and high-volume data.[4,5] Tasks which deal with complex data, such as pattern recognition, are often inefficient when run on von Neumann-type machines. Neuromorphic circuits in silicon have been demonstrated modeling somas[6] and synapses, have shown plasticity[7] and learning, and have been integrated to the system level.[4,8] The results can include much quicker runtimes, smaller size processors, and higher energy efficiency when dealing with such tasks.[5]

Power dissipation is a serious issue if computational neuron models and neuromorphic processors are to be scaled to large network sizes.[9] Biological neurons dissipate about $10^{-11}$ Joules per spike. Neurons made from silicon circuits, however, are over 3 orders of magnitude higher than that at about $2 \times 10^{-8}$ Joules/spike and digital neurons are even higher at $10^{-5}$ Joules/spike.[10] Supercomputers composed of silicon circuits are coming up against a power wall; only a few more generations of supercomputers will be built with their present make-up.[11] Low-power computing solutions such as neuromorphic computing, reversible computing and approximate computing are expected to play important roles moving forward.[11]

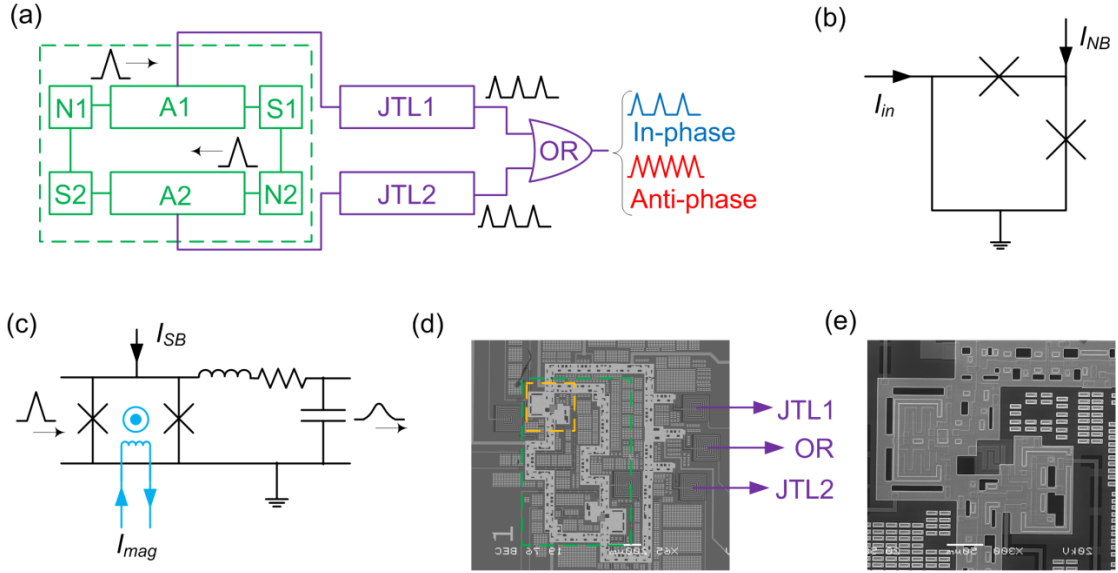

*Figure 1: Experimental details: (a) Block-diagram of the experiment. The first soma-axon-synapse (N1-A1-S1) combination is coupled to the second (N2-A2-S2); this mutually coupled loop (marked by the green dashed line) falls into a synchronized state. A copy of the pulses are coupled out onto JTLs (JTL1 and JTL2) and merged together in an OR gate. The in-phase state will fire at the same frequency as the neurons; the anti-phase state will fire at twice the frequency. (b) Circuit diagram of the two somas, N1 and N2. The X's indicate Josephson junctions. Two currents, $I_{NB}$ and $I_{in}$, bias the soma. (c) Circuit diagram of the synapse. A two-junction loop (DC SQUID) modifies the amplitude of the pulse coming in from the axon. Two currents ($I_{SB}$ and $I_{mag}$) control the amount of modulation. Following the SQUID, an LRC-filter smoothes the pulse; this converts it into a synaptic current. (d) Optical microscope picture of the circuit. The mutually-coupled loop is indicated by the green box, and the S2-N1 combination is indicated by the yellow box. The outputs voltages for the two JTLs and the OR gate are indicated; two other output voltages (not indicated) are taken from the two axons. (e) A zoom-in on the S2-N1 combination.*

In this study we present experimental results on a model system of analog neurons made from low-power, high-speed superconducting Josephson junctions. Two mutually-coupled Josephson junction neurons (JJ neurons) are configured to fire repeatedly, and they quickly transition into a synchronized state. Their synchronization is predominantly of two types: in-phase or anti-phase. Varying the synaptic strength and/or time delay of the connecting synapses causes behavior called a *phase-flip bifurcation*,[12--15] which toggles the system from in-phase to anti-phase or vice-versa. The bifurcation diagrams are measured with excellent agreement with circuit simulations, and are acquired in a fraction of the time to simulate them. The neurons fire at a rate of about 25 GHz and with a power dissipation of about $10^{-17}$ J/spike, easily the highest and lowest numbers, respectively, for analog neurons to date. The speed of the pulses requires indirect detection of the pulses using additional circuitry. We use superconducting Rapid-Single Flux Quantum (RSFQ) circuitry in conjunction with the analog neurons to determine the state of synchronization in real time.

This study provides a proof-of-concept demonstration of the potential of Josephson junction neurons (JJ neurons) and their associated circuitry to model somas, axons, adjustable synapses, and detection of activity states. The JJ neuron toolbox provides possibilities for both neuromorphic computing and fast computational models at high speed and low power.

Methods

A schematic of the circuit is shown in Figure 1a. A JJ neuron[3] (N1), with its full circuit diagram shown in Figure 1b, acts as the first soma. A standard Josephson transmission line (JTL) 20 junctions long acts as the axon (A1), carrying away the pulse from the soma. The axon ends in a synapse (S1), shown in figure 1c. It is formed by a Superconducting Quantum Interference Device (SQUID), similar to earlier work;[16] here we also add an RLC filter. The SQUID changes the amplitude and delay of the action potential by an amount dependent on the flux through the loop, while the filter smoothes and stretches the pulse in time, converting it to a synaptic current. This synaptic current flows to the second soma-axon-synapse (N2-A2-S2) combination, which in turn couples its output back to the first. The whole circuit fires in the range of ~ 25 – 50 GHz, depending on the circuit parameters.

This mutually-coupled loop sometimes synchronizes in one of two synchronized states, in-phase or anti-phase. To detect the synchronized state of the system, a copy of each pulse is split off of its axon, coupled onto separate transmission lines (JTL1 and JTL2) and then merged together in an RSFQ merger (OR gate). The spiking frequency of this OR gate is compared to that of the individual axons. For the in-phase state, the OR gate will fire at the same rate as the neurons; for the anti-phase state, it will fire at twice the frequency of the neurons (Figure 1a). Note that this frequency doubling is for detection purposes only; the base frequency of the neurons do not change.

A picture of the chip, fabricated by Hypres, Inc., is shown in Figures 1d and 1e. The circuit requires seven currents for bias and control. Currents in the mA range for the two halves of the circuit come from a single source and are identical. Two currents (not shown) feed the axons and the JTLs ($I_{axon}$ and $I_{JTL}$). Two currents feed each of the JJ neurons, the neuron bias current and the neuron input current ($I_{NB}$ and $I_{in}$). The merger is biased with a current $I_{OR}$ (not shown). Finally, the synapses involve two control currents: the magnet current ($I_{mag}$) and the synapse bias current ($I_{SB}$). The magnet current is an on-chip current which coupled flux into the SQUID loop, while the synapse bias current provides bias to each of the SQUID Josephson junctions.

The state of the circuit is sensed by measuring the firing frequencies at five different points. Direct Current (DC) voltages are measured at probe junctions on the two axons and the two JTLs, and at the output junction of the OR gate. DC voltages are proportional to frequency via the Josephson relation, 2.07 μV/GHz. The voltages were amplified by a factor of 500, giving a conversion of 1.035 mV/GHz (conveniently about 1mV/GHz).

Experiments were performed at 4.2 K in a $^3$He cryostat from Oxford Instruments. The bias lines were heavily filtered with low-pass filters and powder filters with discoidal capacitors.[17] The currents were provided by custom-made, single-ended current supplies from precision voltage regulators. Voltages were amplified by Analog Devices amplifiers, digitized by a National Instruments Analog-to-Digital converter and saved on disk.

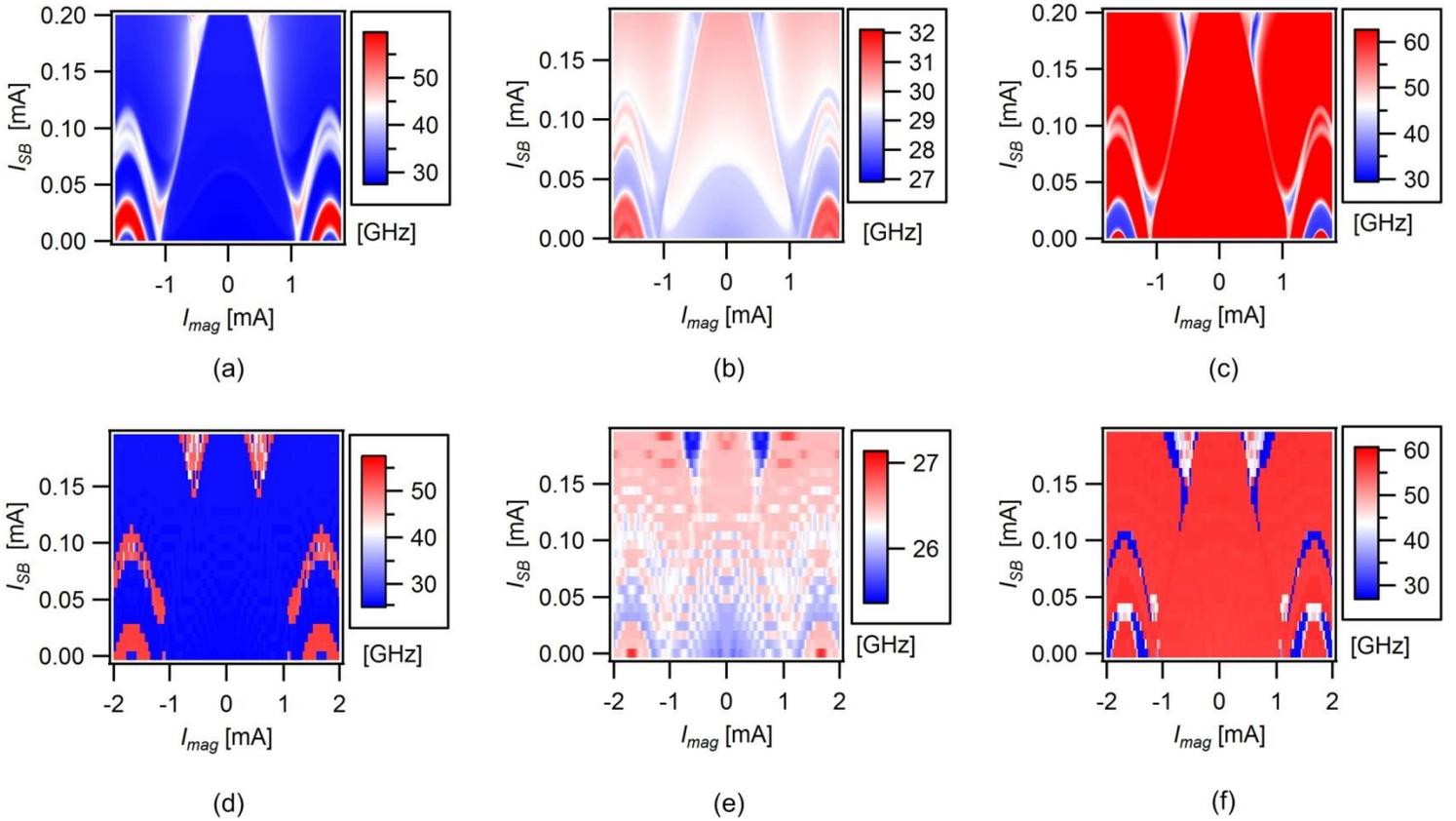

*Figure 2: Experimental Results: (a) Bifurcation diagram. Plotted is the frequency of the OR-gate in GHz as a function of the two synaptic parameters, $I_{SB}$ and $I_{mag}$. Blue regions are at about 29 GHz and represent the in-phase state; red regions are at about 62 GHz and represent the anti-phase state. (b) Frequency of the firing of the neurons, measured at the axons and the JTL probe junctions. (c) Bifurcation diagram with an increased value of $I_{NB}$. The base frequency of the neurons increase, causing a "reversal", where in-phase becomes anti-phase and vice-versa. (d)-(f) Simulated versions of (a)-(c).*

To observe the phase-flip bifurcation, the experimental currents were chosen such that the measured frequency of the two axons and the two JTLs was about 27 GHz, all at the same frequency within the measurement uncertainty (±1 GHz, due to drifting amplifier offsets). The frequencies synchronize if the bias currents are adjusted carefully. With the circuit biased in such a way, we then recorded the voltage of the OR gate as a function of the two synaptic currents ($I_{SB}$ and $I_{mag}$) and plotted a 2-D color plot to form the bifurcation diagram for the system. Typical sweeps were 1000 bias currents by 1000 magnet currents ($10^6$ total points) and took about 17 min., limited by averaging and acquisition time.

Results

In our bifurcation diagrams, we were able to identify several different states of the coupled system. At low synapse bias currents ($I_{SB}$), the neurons were synchronized; here was saw in-phase states, anti-phase states, and bi-stability states (where the system toggles back and forth between in-phase and anti-phase over long timescales). The OR-gate detector records the in-phase states at the same frequency as the neurons, the anti-phase states at about twice the neuron frequency, and the bi-stability states at about 1.5 times the neuron frequency. At high synapse bias currents ($I_{SB}$), the circuit shows another state, a de-synchronized state. In this state, the neurons no longer fire at the same frequency.

We can observe this through the standard deviation of the frequencies measured at the two axons and the two JTLs. We first discuss the synchronized state and the phase-flip bifurcation, and then discuss the de-synchronized state.

Figure 2 shows the bifurcation diagram for parameters where the two neurons are always synchronized. The horizontal axis shows $I_{mag}$ which adjusts the magnetic field inside the synapse's SQUID. The bifurcation diagram has symmetry about zero magnet current, since standard SQUID response to a magnetic field is symmetric. In Figure 2a, we show the frequency of the OR gate in a blue-red color plot with blue at about 29 GHz and red at about 58 GHz; blue corresponds to an in-phase state, red to an anti-phase state, and white to a bi-stability state. In Figure 2b, we show the frequency of the neuron itself, measured at the probe junctions along the axon, also in a blue-red color plot but at a reduced frequency range (27-32 GHz). The firing frequency changes by only about 15 percent over the whole range of magnet and bias current, demonstrating that the frequency changes in the OR gate in Figure 2a are clearly due to phase changes, and not because the whole circuit is oscillating at a higher frequency. Figure 2b also shows a shift in frequency of the underlying neurons at the phase-flip bifurcation, with the in-phase state near 29 GHz and the anti-phase state near 31 GHz.

Phase-flip bifurcations are expected in any coupled-oscillator system when the coupling strength and/or the time delay are varied.[13] The characteristic features of a phase-flip bifurcation are: (1) a phase change between the two oscillators, (2) a change in the firing frequency between the two synchronization states, and (3) a discontinuity the Lyaponov exponent. The first two observations are clearly seen in Figures 2a and 2b, respectively; we show the discontinuity in the Lyaponov exponent in the appendix. This demonstration of a phase flip-bifurcation in an analog neuron system is at an extremely high frequency compared to previous work.[18,19] In addition, the bifurcation occurs as a function of three different parameters: $I_{mag}$, $I_{SB}$ and $I_{NB}$, more than we had expected. This suggests that phase-flip bifurcations may be very common in neuron networks, an important observation since synchronization is used[20] by the brain in many different areas of sensory and information processing.

The entire circuit was also simulated with WR-SPICE from Whitely Research. Since the bifurcation diagram varies two parameters, the remaining five control currents ($I_{NB}$, $I_{axon}$, $I_{JTL}$, $I_{in}$ and $I_{or}$) and another parameter, the chip critical current density, all need to be chosen as input to the simulation. Even our small measurement uncertainty led to a complex fitting procedure in this six-dimensional space. For the fitting shown in Figure 2, all fitting parameters are within 8% of the experimental parameters. Figures 2d and 2e show the simulated neuron frequencies and OR-gate frequencies, in direct comparison with Figures 2a and 2b. The in-phase, anti-phase and bistable states appear at mostly similar values of $I_{SB}$ and $I_{mag}$, showing we have achieved good agreement over this entire two-dimensional parameter range.

In Figure 2c we show a similarly obtained bifurcation diagram with a larger value of the neuron bias current ($I_{NB}$), which increases the firing rate of the neurons to about 31 GHz for the in-phase state. Because this alters the time in which the pulse arrives at the postsynaptic soma, with the appropriate choice of bias current the synchronization state can be flipped. With the chosen parameters, the entire bifurcation diagram is in fact "inverted", where in-phase becomes anti-phase and anti-phase becomes in-phase (red becomes blue and blue becomes red). The circuit simulation, using the same parameters including the change in $I_{NB}$, is shown in figure 2f. The simulation is able to reproduce this inversion as well.

The synaptic parameters we have varied in Figure 2, $I_{SB}$ and $I_{mag}$, are specific to our Josephson system. In biological synapses, the typical synaptic parameters are synaptic strength and delay. In figure 3 we make the conversion from our parameters to the biological ones. Figure 3a shows the delay (from soma to synapse) in picoseconds and figure 3b shows the strength (amplitude of the pulse coming out of the synapse in mV), both as a function of $I_{SB}$ and $I_{mag}$, for the biasing parameters used in Figure 2d and 2e. Using this information allows us to reconstruct the bifurcation diagram of figure 2d in delay-amplitude space. We show this in figure 3c, where we plot the frequency of the OR-gate (indicator of in-phase or anti-phase) as a function of amplitude and delay. The bifurcation is primarily in the delay direction, but figure 3c shows some non-trivial dependence on amplitude as well. Synchronization does not appear to be significant until a threshold synaptic strength of 0.3 mV is reached. This suggests a threshold-like behavior in which synchronization occurs only for sufficiently large synaptic strength. Experiments in biological systems to describe such a threshold would be interesting.

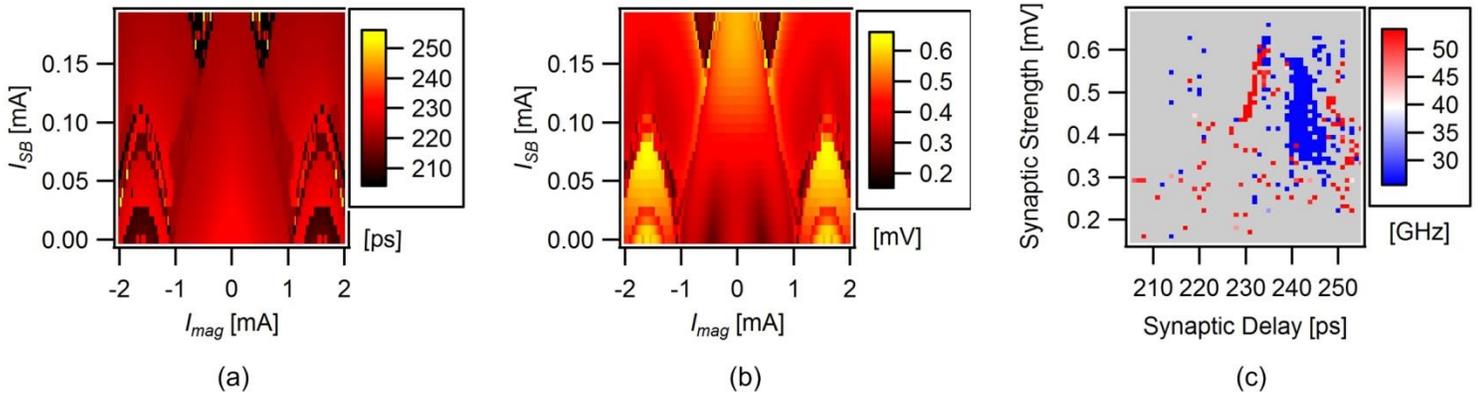

*Figure 3: Synaptic strength and delay. (a) Synaptic delay. Plotted is the delay from soma to synapse in picoseconds (ps) as a function of the two synaptic parameters, $I_{SB}$ and $I_{mag}$. The range of parameters is identical to figure 2. (b) Synaptic strength. Plotted is the amplitude of the synaptic pulse in mV versus $I_{SB}$ and $I_{mag}$, again over the same range as figure 2. (c) Bifurcation diagram plotted in the strength-delay space. The blue points are the in-phase state and the red points are the anti-phase state, as before.*

As already mentioned, for the range of parameters shown in the bifurcation diagrams in figure 2 the circuit is always in synchronization. However, if we go to larger synapse bias currents ($I_{SB}$), the circuit falls out of synchronization. Figure 4 shows the standard deviation of the firing frequency at the four different outputs in the circuit (one on each axon, one on each JTL). In the range used in the diagrams in figure 2 (dashed box), the standard deviation is less than 1 GHz, within our measurement uncertainty, implying full-circuit synchronization. However, in the dark regions outside of the dashed box, the standard deviation gets as high as 10 GHz. This is a distinct state of the circuit, a de-synchronized state. The implication of this is discussed below.

Discussion

The fact that we have simulated neuron behaviors on the picosecond timescale with biologically-realistic neurons allows one to consider applications in the study of long-term neuron dynamics. Although we have only studied two neurons here, it is straightforward to add more neurons since they are made lithographically. Consider, for example, the study of epilepsy, where large-scale synchronization of thousands to millions of neurons causes epileptic seizures. Existing biophysical network models run on digital computers[21] have already made progress in understanding aspects of epilepsy, such as the shift[22] from dominant excitation to dominant inhibition, the modeling[23] of absence seizures, and the mechanisms[24] relating fast oscillations to seizure onset. Digital simulations, however, are hampered by the limited ability of computers to model collective neuron dynamics in large networks over long time-scales. An approach using superconducting ICs with analog neurons such as demonstrated here would allow the same models to be studied but with computational times thousands of times faster, permitting a vastly greater exploration of parameter space.

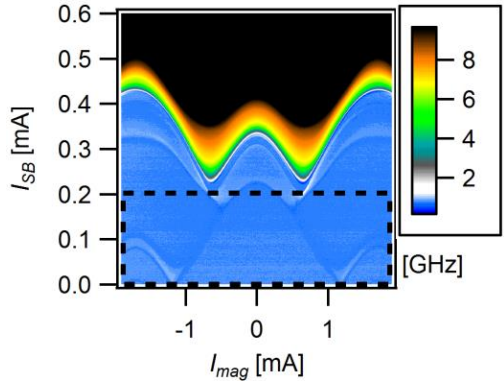

*Figure 4: Standard deviation of the time-averaged firing frequency at four points in the circuit, one on each axon and one on each JTL, plotted as a function of $I_{SB}$ and $I_{mag}$. The dashed black box represents the region of figure two, where the circuit is in full synchronization. At higher values of $I_{SB}$, however, the circuit becomes unsynchronized.*

The data in Figure 4 gives a preview of how one might study epilepsy in larger neural networks in the future. In a seizure, the firing frequencies become essentially the same; that is the standard deviation of the firing frequencies decreases. So, a measurement of the standard deviation across neurons of the time-averaged firing frequency is an indicator of seizure-like behavior. Figure 4 shows the standard deviation across four points in our circuit as an indicator of seizure-like behavior in our two neuron system. With more neurons in the circuit, we would simply take the standard deviation over more neurons. The dark, de-synchronized regions correspond to parameter regions where behavior is seizure-like, while the light blue synchronized regions correspond to parameter regions where behavior is not seizure-like. The digital calculation of the information in figure 4 takes about 60 hours with 1000 times fewer points; our JJ neuron simulation took 51 minutes, a factor of over 70,000 times faster per data point in an un-optimized system. Importantly, adding more neurons would not add to the time needed for these JJ neuron simulations.

These results demonstrate that we have been able to construct an analog system where we can control its behavior (synchronization) with off-chip currents. In the area of neuromorphic computing, we have demonstrated spiking neurons, axons, adjustable synapses, and collective-state detection for a system dissipating $10^{-17}$ J per pulse and firing at tens of GHz. Future experiments are expected to show both long-term plasticity (LTP) and long-term depression (LTD) as well as spike timing dependent plasticity (STDP). LTP can be implemented by using the pulsed output of our state detector to vary one of the synaptic parameters ($I_{SB}$ and $I_{mag}$) in a manner consistent with a learning[7] rule. A "forgetting" resistor can be put into the SQUID loop such that the current state has LTD. For STDP, one could use an RS-Flip Flop[25] as the state detector to ensure that synaptic increases only occur if the presynaptic neuron fires before the postsynaptic neuron and not the other way around. With the toolbox of RSFQ comparison circuits, variable[26] couplers, and superconducting transmission lines and filters already demonstrated by the superconducting digital community and the superconducting quantum computing community,[27] these should be straightforward next steps, enabling a fast, low-power learning network.

In conclusion, we have demonstrated the operation of a superconducting neuromorphic circuit via a synchronized oscillator behavior known as a phase-flip bifurcation. With the high speed and low power dissipation of superconducting circuits, vast improvements in neuronal computation are possible.

We thank A. Kadin, E. Olson, J. Meyers and B. Hansen for useful discussions and T. Griffin and E. LaGorga for simulation assistance. This work was supported by the Picker Interdisciplinary Science Institute.

Appendix

In this appendix we support the claim that the bifurcation we see in the experimental and simulation data is a phase-flip bifurcation.[13] We also reproduce the basic character of the bifurcation using a simplified model. The simplified model is easier to analyze and we use it to explore behavior details.

We proceed to describe the simplified model of two mutually coupled JJ Neurons. We then present simulated time profiles of in-phase and anti-phase solutions, a bifurcation diagram of solution character versus delay and synaptic strength comparable to Figure 3c, and the three characteristic behaviors of solutions across a phase-flip bifurcation. These include an abrupt change in the phase (in- vs anti-) character of the solution, a jump in period of oscillation at the bifurcation point, and a discontinuity in the Lyapunov exponents at the bifurcation. Together this supports the claim of a phase-flip bifurcation by reproducing the behavior using a simplified model in which analysis of the details of the bifurcation is more accessible.

The simplification in our model comes from replacing the axon portions of the circuit with a delay to represent the time it takes for the signal to move down the axon to the synapse. The circuit diagram in Figure 5 shows how the delay replaces a portion of the circuit. The resulting system of equations are delay differential equations (DDEs) with the synapse variables dependent on a delayed version of the variables in the soma. DDEs are equivalent to infinite dimensional systems of ordinary differential equations (ODEs), so our simplification is arguably not simple. But the interactions of equations describing the axon can be subtle and the resulting time of transmission is hard to adjust explicitly. In the DDE model the transmission delay is an explicit parameter representing the time of transmission along the axon. We can adjust this transmission delay parameter directly. While the total delay between soma excitation and synapse excitation involves the sum of the transmission delay and the excitation time of the synapse, our delay parameter allows us to control the total delay quite effectively. This is the sense in which the model is simple.

Replacing the axon with a delay, the remaining circuit elements represent the soma and synapse portions of the system. These portions are modeled using the electric circuit equations described in Crotty et al.[3] The relevant variables are $\varphi_{pi}$, $\varphi_{ci}$, $v_{pi}$, $v_{ci}$, $v_{outi}$, $f_{outi}$, $i_{12}$ where subscript $i \in \{1,2\}$ denotes which neuron is described. The $\varphi$ variables describe the phase jump across the junctions, the $v$ variables are voltages across the junctions and synapse capacitor (out) and $i_{12}$ is the current through the synapse resistor toward the next neuron. There are 14 variables and 14 equations:

$$\dot{\varphi}_{p1} = v_{p1} \tag{1}$$

$$\dot{v}_{p1} = -\Gamma v_{p1} + \sin(\varphi_{p1}) + \Lambda_s\, i_{12} - \lambda\,(\varphi_{p1} + \varphi_{c1}) + (1 - \Lambda_p) i_b \tag{2}$$

$$\dot{\varphi}_{p2} = v_{p2} \tag{3}$$

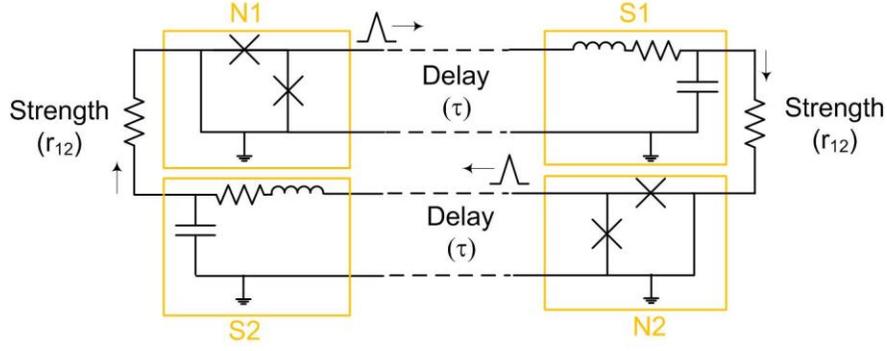

*Figure 5: Simplified circuit of figure 1a. The neurons connect to the synapses with an adjustable delay τ. The synapses connect back to the neurons with a resistor $r_{12}$; changing its value adjusts the strength of the synapse.*

$$\dot{v}_{c1} = -\Gamma v_{c1} + \sin(\varphi_{c1}) + \Lambda_s\, i_{12} - \lambda\,(\varphi_{p1} + \varphi_{c1}) - i_b \Lambda_p \tag{4}$$

$$\dot{v}_{out1} = f_{out1} \tag{5}$$

$$\dot{f}_{out1} = -\Omega_o Q\, f_{out} - i_{12}\frac{\Omega_o^3 Q \Lambda_{syn}}{\lambda} - \frac{\Omega_o^2 \Lambda_{syn}}{\lambda}(\dot{i}_{12}) + \Omega_o^2\,[v_{p2}(t-\tau) - v_{out1}] \tag{6}$$

$$(\dot{i}_{12}) = \frac{\lambda}{\Lambda_s(1-\Lambda_s)}\left[v_{out1} - \frac{r_{12}}{\Gamma} i_{12} - \lambda(\varphi_{p1} + \varphi_{c1})\right] \tag{7}$$

$$\dot{\varphi}_{p2} = v_{p2} \tag{8}$$

$$\dot{v}_{p2} = -\Gamma v_{p2} + \sin(\varphi_{p2}) + \Lambda_s\, i_{21} - \lambda\,(\varphi_{p2} + \varphi_{c2}) + (1-\Lambda_p)i_b \tag{9}$$

$$\dot{\varphi}_{p2} = v_{p2} \tag{10}$$

$$\dot{v}_{c2} = -\Gamma v_{c2} + \sin(\phi_{c2}) + \Lambda_s\, i_{21} - \lambda\,(\varphi_{p2} + \varphi_{c2}) - i_b \Lambda_p \tag{11}$$

$$\dot{v}_{out2} = f_{out2} \tag{12}$$

$$\dot{f}_{out2} = -\Omega_o Q\, f_{out2} - i_{21}\frac{\Omega_o^3 Q \Lambda_{syn}}{\lambda} - \frac{\Omega_o^2 \Lambda_{syn}}{\lambda}(\dot{i}_{21}) + \Omega_o^2\,[v_{p1}(t-\tau) - v_{out2}] \tag{13}$$

$$(\dot{i}_{21}) = \frac{\lambda}{\Lambda_s(1-\Lambda_s)}\left[v_{out2} - \frac{r_{21}}{\Gamma} i_{21} - \lambda(\phi_{p2} + \phi_{c2})\right] \tag{14}$$

Parameters values include $\eta = 1$, $\Gamma = 1.55$, $\lambda = 0.13$, $\Lambda_s = 0.487$, $\Lambda_p = 0.482$, $i_{12} = 2.2$, $\Lambda_{syn} = 0.3$, $\Omega_o = 1$, $Q = 0.05$. The parameters $r_{12} = r_{21}$ represent the strength of the synapse and our control parameter is $\tau$ representing the transmission delay.

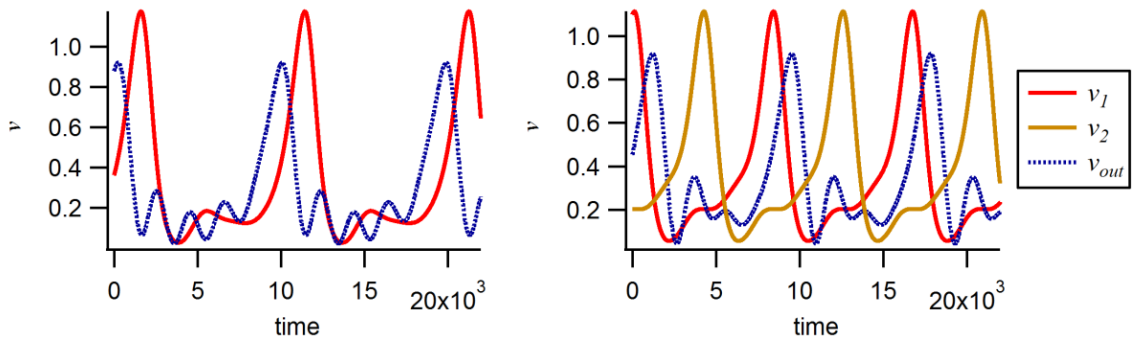

*Figure 6: Time profile for the in-phase state (left) and the anti-phase state (right). Shown are the two neuron voltages ($v_1$, $v_2$) and the output voltage of the first synapse ($v_{out}$). Note that in the symmetric case $v_1$ and $v_2$ are exactly on top of each other, so that $v_2$ is not visible. The synapse pulse switches from before $v_1$ to after $v_1$ across the bifurcation.*

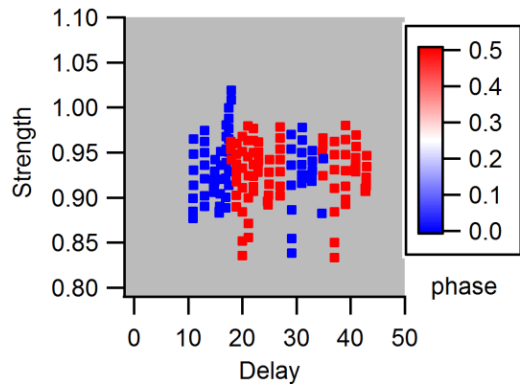

*Figure 7: Bifurcation diagram in delay-strength space. Blue points represent in-phase while red points represent anti-phase. The bifurcation occurs primarily in the delay direction, with some non-trivial dependence on strength. This has similarities with Figure 3c in the main text.*

Solving these equations, using standard DDE methods, results in time profiles shown in Figure 6. Figure 6a shows the in-phase case when $r_{12} = 1.4$ and $\tau = 16.0$ while 6b shows the anti-phase case when $r_{12} = 1.4$ and $\tau = 17.0$. Notice that the in-phase case displays both neuron's voltages directly on top of each other, while the anti-phase case has them shifted by a half period relative to each other. The relative timing of the synapse pulse and the second neuron pulse indicates a possible mechanism for the bifurcation. For in-phase solutions, the synapse pulses just before the neuron it connects to; presumably it helps the neuron to fire. For the anti-phase solution, the synapse pulses just after the neuron it connects to so the synapse doesn't lead to immediate firing of the neuron. It does shorten the overall period, however; presumably the early excitation moves the neuron through the slow build-up phase of the oscillation to a state closer to its firing threshold. In any case, the bifurcation appears to occur when the delay between soma and synapse firing crosses a multiple of the period.

Varying both $r_{12}$ and $\tau$ while tracking the character and period of the oscillation produces the bifurcation diagram shown in Figure 7. The characteristic stripe pattern of in-phase and anti-phase solutions resembles Figure 3c in the paper for simulations of the more complex system. There are regions of bi-stability near the boundaries of the colored regions. We suspect that such bi-stability leads to bands of white in the experimental data since that data is averaged over many observations yielding a result between the pure in-phase or pure anti-phase result. Crossing the boundary involves a phase-flip bifurcation.

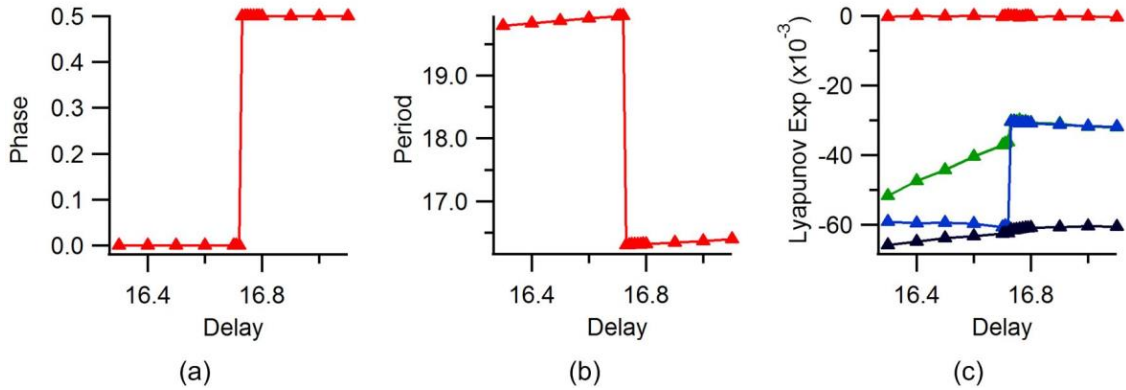

*Figure 8: Three characteristics of a phase-flip bifurcation. (a) Phase versus delay. The phase flips from in-phase (0) to anti-phase (0.5) at the bifurcation. (b) Period versus delay. The period drops about 15% at the bifurcation point. (c) Lyapuonov exponents versus delay. The highest exponent (red) is zero, representing the perturbations along the periodic solution. The second-, third- and fourth-highest exponents are shown in blue, green and black, respectively. The second and third undergo a discontinuity at the bifurcation point. Note that the green curve is directly under the blue curve after the bifurcation.*

The three characteristics of a phase-flip bifurcation as described by Prasad et al. are a change in the phase character of the time profile, a jump in period/frequency of oscillation, and a discontinuity in Lyapunov exponent at the bifurcation point.[13] Figure 8 shows all of these characteristics. Fixing $r_{12} = 1.4$ and varying $\tau$ through the bifurcation point provides the time profiles shown in Figure 6. In addition the period jumps from 19.870 to 16.367, as shown in Figure 8(b). The leading Lyapunov exponents are displayed in Figure 8(c). These are computed using the DDE methods from Farmer et al.[28] Notice the zero Lyapunov exponent present for all periodic orbits. Not all exponents have a jump at the bifurcation point, but e.g. the second- and fourth-largest do. Together these findings indicate a phase-flip bifurcation that occurs when delay changes and in some circumstances when synaptic strength changes (as measured by $r_{12}$). This strongly suggests that the same bifurcation is occurring in the full simulated system as well as in the experiments.